\documentstyle[12pt]{article}
\begin{document}

\baselineskip=24pt

\begin{titlepage}
\begin{flushright}
PVAMUHEP-92-2 \\
FSUHEP-921126
\end{flushright}

\vskip.15in

\begin{center}
{\Large A $Z'$ Model Which May Be Relevant to the New LEP Events}\\
(revised)

\bigskip

Dan-di Wu \\
High Energy Physics, P. O. Box 355, Prairie View A \& M University \\
Prairie View, Texas 77446 USA \\
and \\
Superconducting Super Collider Laboratory \\
2550 Beckleymeade Avenue, Dallas, Texas 75237 USA \\

\medskip

and

\medskip

Chung Kao \\
Department of Physics, B-159, Florida State University \\
Tallahassee, Florida 32306 USA\\
\end{center}

\vskip .35in

\begin{center}
{\bf Abstract}
\end{center}

A multi-Higgs model with an extra neutral gauge boson ($Z'$)
is introduced.
One scalar Higgs boson ($H_2$) in this model decays dominantly
into a photon pair.
The $Z'$ decay to $\mu^+\mu^-$ gets a much larger
branching ratio than the $Z$ decay to this channel.
The $Z Z' H_2$ vertex provides  a final state from $Z$ decay
resembling the new $l^+l^-\gamma\gamma$ events at the LEP.
Other promising phenomenology, such as $Z \rightarrow l\bar l Z'$
is also discussed.

\medskip


\end{titlepage}
\newpage

A simple way to extend the standard model of the electroweak theory
\cite{STEVE} (SM) is to add an extra $U(1)'$ gauge boson.
These so called extra $Z$ models \cite{NEWU1} may be originated
from all kinds of motivations and predict different phenomenology
to be tested in high energy experiments.

He-Joshi-Lew-Volkas \cite{HJLV}(HJLV) have discussed a set of three simplest
$Z'$ models. The common feature of these models is that the $Z'$,
the gauge boson of an extra $U(1)'$ gauge symmetry, only couples
to leptons and neutrinos, so the leptonic decay of $Z'$ is greatly enhanced.
This feature makes these models seem relevant to the new LEP events.
Indeed, recently at the DPF92 Meeting, S.~Ting \cite{SAM} reported 4 peculiar
$l \bar l \gamma \gamma$ events from L3, three of them are
$\mu \bar{\mu} \gamma \gamma$ and one $e \bar e \gamma \gamma$.
There seems to be a two photon invariant mass clustering around 59.7 GeV
within two standard deviations.
DELPHI \cite{DELPHI} and ALEPH \cite{ALEPH} also have similar events,
2 from each. In addition, there are about ten $l \bar l \gamma \gamma$
events from each group that spread over a large range of $\gamma\gamma$
invariant masses.
Although it is still too early to conclude anything from these peculiar events,
it is interesting to see how one of the HJLV models may be relevant to
these events. We shall see that a modified HJLV model is particularly
of interest in this context.

The modification of the HJLV models we shall introduce is to allow a $Z_1-Z_2$
mixing, where $Z_1$ and $Z_2$ are respectively the SM $Z$ boson and the gauge
boson of the extra $U(1)'$ symmetry with mass eigenstates $Z$ and $Z'$
and we identify $Z$ as the one actually discovered at LEP.
Because of this mixing, a $Z Z' H_2$ vertex is present in the modified models.
 The models then allow the following process to happen at LEP
\begin{equation}
Z \rightarrow Z'^* H_2 \rightarrow l \bar l H_2,
\end{equation}
where $H_2$ is a scalar component of the second Higgs doublet which, we assume,
does not couple to any fermions\cite{THDM,BDHR} because it has a nonzero
$U(1)'$ quantum number.
$H_2$ mainly decays into two photons. Its width is about 1 keV (see later),
if its mass is about 60 GeV and it does not significantly mix with $H_1$,
the SM Higgs boson. $H_2$ therefore does not decay into fermions through tree
diagrams when the small mixing is neglected. It may still decay into fermion
pairs via $W^\pm$, $Z$ and charged Higgs loops. However our calculation shows
that the total fermionic branching ratio of $H_2$ is small.
The qualitative reason of this is that
the $H_2 \rightarrow 2 \gamma$ loops involve derivative couplings
while $H_2 \rightarrow f \bar f$ loops have only direct couplings.

The first candidate which come to our consideration is model I of HJLV.
In this model the $U(1)'$ quantum numbers of $e$ and $\mu$ families sum
to zero. In other words, the $U(1)'$ gauge boson $Z_2$ only couples to $e$,
$\mu$ and their neutrinos. Because there are only left handed neutrinos,
as are in the SM, the $Z_2$ decay branching ratio satisfy $e\bar e : \mu
\bar \mu : (\nu_e \bar \nu_e + \nu_\mu \bar \nu_\mu) = 1:1:1$. In any case
$Z_2$ has a much larger $e,\mu$ branching ratio than $Z_1$ does. This feature
allows the model to give an explanation as of why $q \bar q \gamma \gamma$
is not observed and, perhaps due to reasonable statistical fluctuations,
$\nu \bar \nu \gamma\gamma$ has still been missing \cite{SAM,DELPHI,ALEPH}.
Unfortunately, this model is strongly restricted by available
$\Delta R(e^+e^- \rightarrow \mu^+ \mu^-)$ measurements\cite {HIKASA,NEWZ}.
The coupling
constants of $Z'$ to leptons have to be so small so that the calculated
branching ratio of $Z \rightarrow Z^{'*} ( \rightarrow l \bar l) H_2$
can only reach $few \times 10^{-8}$, which is too small to account
for the L3 events.

The next candidate, which looks more promising is model III of HJLV.
In this model, the $U(1)'$ quantum numbers of the $\mu$ and $\tau$ families
sum to zero. A weakness of this model is that, in addition to blaming the
lack of $\nu \bar \nu \gamma \gamma$ in the L3 experiment to statistical
reasons
we leave the observed $e^+e^- \gamma \gamma$ unexplained.
It is worth commenting at this point that most of these 8 events
(except one or two $\mu^+ \mu^- \gamma \gamma $ events) are, still peculiar,
but somehow  one can always find a $\gamma - l$ pair whose total energy
is almost the beam energy. The original version of model III is only
very loosely restricted by $(g-2)_\mu$ loop diagram \cite{HJLV}
$$ m_{Z'}>100 y'g' \ {\rm GeV},$$
where $g'$ is the $U(1)'$ gauge coupling constant
and $y'$ the $U(1)'$ quantum number of the fermions.
The modified model will be restricted by the LEP experiments.
Now let us describe this model in some detail.

The mass of $Z_2$ comes from two sources: a Higgs singlet $S$  which
contributes
the main part of the $Z_2$ mass and an extra Higgs doublet $\phi_2$ which
contributes masses to $Z_1$, $Z_2$ and their mixing.
The relevant Higgs-gauge couplings are
\begin{eqnarray}
{\cal L}_{HZZ}& = &
\frac{1}{4} \frac{g_2^2}{1-x} Z_1^2 ( \frac{v_1+H_1}{\sqrt{2}} )^2 \nonumber \\
              &   &
+\frac{1}{4} ( \frac{g_2}{\sqrt{1-x}} Z_1 - g' Z_2 )^2
 ( \frac{v_2+H_2}{\sqrt{2}} )^2
+g'^2 Z_2^2 ( \frac{v_3+H_3}{\sqrt{2}} )^2.
\end{eqnarray}
where $x= \sin^2 \theta_W$ with $\theta_W$ being the Weinberg angle
\cite{STEVE}.  $g_2$ is the $SU(2)$ gauge coupling constant, $g_2=e/\sqrt{x}$,
and $e$ is the electric charge of the proton.
Here we assume that the $U(1)'$ quantum numbers for the Higgs doublets
$\phi_1, \phi_2$ and the singlet $S$ are respectively 0, 1/2 and 1 and
\begin{equation}
< \phi_1 > = \frac {v_1}{\sqrt{2}}, \,\,
< \phi_2> = \frac{v_2}{\sqrt{2}}, \,\,
< S > = \frac{v_3}{\sqrt{2}},
\end{equation}
The mass matrix in the $Z_1 -Z_2$ basis is
\begin{eqnarray}
\left(
\begin{array}{clcr}
m_{Z_1}^2              & -\frac{1}{4} \frac{g_2}{\sqrt{1-x}} g'v_2^2 \\
-\frac{1}{4} \frac{g_2}{\sqrt{1-x}} g'v_2^2  &  m_{Z_2}^2
\end{array}
\right)
\end{eqnarray}
with
\begin{eqnarray}
m_{Z_1}^2 & = &\frac{1}{4} \frac{g_2^2}{1-x} v^2, \,\,
m_{Z_2}^2  = {g'}^2 {v_3}^2 + \frac{1}{4} {g'}^2 v^2 \sin^2 \beta, \nonumber \\
\sin \beta & =& \frac{v_2}{v} , \,\,
v^2        = v_1^2 + v_2^2.
\end{eqnarray}
Let $Z$ and $Z'$ be the mass eigenstates with
\begin{equation}
Z = Z_1 \cos{\alpha_Z} - Z_2 \sin {\alpha_Z},\,\,
Z'=Z_1 \sin {\alpha_Z} + Z_2 \cos {\alpha_Z},
\end{equation}
after diagonalizing the mass matrix, we have, for small mixings
\begin{eqnarray}
\tan {\alpha_Z}       & = &
\frac{g'\sqrt{1-x}}{g_2}(\frac{m_Z^2}{m_Z^2-m_{Z'}^2}) \sin^2 \beta,
\nonumber \\
\frac{\Delta m_Z}{m_Z}& = &{m_Z-m_{Z_1} \over m_Z}=
\frac{1}{2} (\frac{m_{Z'}^2-m_{Z}^2}{m_Z^2}) \tan^2 {\alpha_Z}.
\end{eqnarray}

The relevant fermion-gauge sector is, assuming the $U(1)'$ quantum numbers
for the $\mu$'s and $\tau$'s are respectively $-y'$ and $ y'$ with $y' >0$,
\begin{eqnarray}
{\cal L}_{Zff}& = &
Z^{\mu}[(ea_L\cos{\alpha_Z}
+g'y'\sin{\alpha_Z} )\bar{\mu}_L\gamma_\mu \mu_L
+(ea_R\cos{\alpha_Z}
-g'y'\sin{\alpha_Z} )\bar{\mu}_R\gamma_\mu \mu_R \nonumber \\
              &   &
(ea_L\cos{\alpha_Z}
-g'y'\sin{\alpha_Z} )\bar{\tau}_L \gamma_\mu \tau_L
+(ea_R\cos{\alpha_Z}
+g'y'\sin{\alpha_Z} )\bar{\tau}_R\gamma_\mu \tau_R \nonumber \\
              &   &
+(ea_\nu\cos{\alpha_Z}
+g'y'\sin{\alpha_Z})\bar{\nu}_{\mu L}\gamma_\mu \nu_{\mu L}
+(ea_\nu \cos{\alpha_Z}
-g'y'\sin{\alpha_Z})\bar \nu_{\tau L}\gamma_\mu \nu_{\tau L}
]
\end{eqnarray}
and
\begin{eqnarray}
{\cal L}_{Z'ff}& = &
{Z'}^{\mu}[(-g'y'\cos{\alpha_Z}
+ea_L\sin{\alpha_Z}) )
\bar{\mu}_L\gamma_\mu\mu_L
              +( g'y'\cos{\alpha_Z}
+ea_R \sin{\alpha_Z})
\bar{\mu}_R \gamma_\mu \mu_R \nonumber \\
&   & +(g'y'\cos{\alpha_Z}
+ea_L\sin{\alpha_Z}) )
\bar{\tau}_L\gamma_\mu \tau_L
              +( - g'y'\cos{\alpha_Z}
+ea_R \sin{\alpha_Z} )
\bar{\tau}_R \gamma_\mu \tau_R \nonumber \\
             &   & +( - g'y'\cos{\alpha_Z}
+ea_\nu\sin{\alpha_Z} )
\bar{\nu}_{\mu L}\gamma_\mu \nu_{\mu L}
              +( g'y'\cos{\alpha_Z}
+ea_\nu\sin{\alpha_Z} )
\bar{\nu}_{\tau L}\gamma_\mu \nu_{\tau L}]
\end{eqnarray}
with
$$a_L=(x-\frac{1}{2})/\sqrt{x(1-x)},\,\,a_R=\sqrt{x/(1-x)},\,\,
a_\nu=1/2 \sqrt{x(1-x)}.$$
Note that the charged lepton-$Z_2$ coupling here is axial-vector like.

The branching ratio of process (2) is
\begin{equation}
B_{new} =
\frac{1}{288\pi} \frac{\alpha^2}{\sqrt{x(1-x)}}
\tan\alpha_Z (\frac{g'}{e})
[ 0.711\,\frac{\Gamma_{Z'}(m_Z^2-m_{Z'}^2)}
              {\Gamma^Z_{\mu\mu} \Gamma_Z m_{Z'} } ]{\cal I}
\end{equation}
where $\Gamma_{Z'}$ and $\Gamma^Z_{\mu\mu}$ are respectively the total
width of the $Z'$ and the leptonic width of the $Z$.
${\cal I}$ is proportional to the integral of the matrix element
in the phase space and $\alpha = e^2/4\pi $.
\begin{eqnarray}
{\cal I}& = &-\frac{1}{2}(47-y^2-8\delta- \frac{8(y^2-\delta^2)}{z^2})(1-y^2)
             +3(4+2y^2-16\delta+4\delta^2)ln(1/y) \nonumber \\
        &   &+\frac{3(8y^2+4\delta-16\delta^2+4\delta^3)} {\sqrt{y^2-\delta^2}}
              \cos^{-1} [ \frac{2y^2-\delta(1+y^2)}{z^2y} ]
\end{eqnarray}
with
\begin{equation}
y  = \frac{m_{H_2}}{m_Z}, \,\,
z  = \frac{m_{Z'}}{m_Z}, \,\,
\delta = \frac{1}{2}(1+y^2-z^2).
\end{equation}
When $z=1$, this integral becomes the Bj-function \cite{BJ,GUIDE}.
${\cal I}$ in Eq. (11) is only valid for
\begin{equation}
1-y < z < 1+y,\,\,\,\Gamma_{Z'}/m_Z << z-(1-y),
\end{equation}
and it is enhanced quickly when $z$ approaches $1-y$,
$i.$ $e.$, $m_{Z'}$ approaches $m_Z -m_{H2}$.
Although the parameters in the original model III are quite arbitrary, the
modified model is strongly restricted by the available LEP data.
The small inaccuracy in
the leptonic width of the $Z$ and the forward-backward asymmetry of the muon
and
the tau \cite{HIKASA} require $g'y'\sin\alpha_Z/e$ to be less than about 0.01.
The small deviation between estimated and measured $Z$ mass puts a constraint
on $(m_Z^2-m_Z'^2)\tan^2\alpha_Z/m_Z^2$ which must be less than about 0.01.
In order that $B_{new}$ has a reasonable value to be relevant to the L3 events,
all parameters, $\sin\alpha_Z,\, g'y'$ and $m_{Z'}$ have to take the most
optional values. $B_{new}$ will be too small if
$m_{Z'}$ value is larger than 40 GeV because $B_{new} $ is
very sensitive to $m_{Z'}$, see Figure.1.
As an example, setting $(g'y')^2/e^2 =0.01,\,\sin \alpha_Z=0.1$ and $x=0.23$
we find
\begin{eqnarray}
 B_{new} & \sim & 0.60 \times 10^{-7}|1-z^2|{\cal I},\\
B_\mu:B_\tau:B_\nu &=& 50:3:47.
\end{eqnarray}
Setting further $g'=e$ , $m_{H_2}=59.7$ GeV and
$B_{new}=(0.5-1.2)\times 10^{-6}$ we obtain
\begin{equation}
32.5<m_{Z'}<34.0\ {\rm GeV},\,\, \Gamma_{Z'}= 4.2-4.4\ {\rm MeV}.
\end{equation}
A lighter $Z'$ than those in this region can provide a larger
branching ratio of process (2),
however, it will require $Z'$ to be produced on mass shell or almost
on mass shell which seems not coinciding with the L3 events.
The situation with $Z'$ on mass shell while $H_2$ off mass shell is very
unfavorable because the width of $Z'$ is much larger than that of $H_2$,
which is
\cite{GUIDE}
\begin{equation}
\Gamma_{H_2} \simeq 1.6 \sin^2\beta\ {\rm keV}.
\end{equation}
In our above parameterization, $\sin^2 \beta \sim 0.1$.
This process is dominated by the $W^\pm$ loop. The same virtual $W^\pm$ pair
may also transfer into a fermion pair. However its total width is about five
times smaller than this.

It seems that the $Z A H_2$ and the $Z Z H_2$ vertices can also produce
$f \bar f \gamma \gamma$ events,
where $A$ is the physical pseudoscalar Higgs boson.
Both modes can be made comfortable
by introducing a large $\sin^2 \beta$ and smaller $g'$ (or neglecting
$Z'$ completely\cite {BDHR}). However the first one will be
dominated by heavy fermions such as $b \bar b$. The second one, as we know,
will produce too many $q\bar q$ and $\nu \bar \nu$. These modes look more
unlikely to be relevant.

A $Z'$ so light can be produced on-shell in some processes, in particular
\begin{equation}
Z \rightarrow l\bar l \rightarrow l \bar l Z'.
\end{equation}
where one of the lepton(or anti-lepton) in the intermediate step
is off mass shell.
With the same parameterization to obtain Eq. (16), we find that this mode
has a reasonably large total branching ratio which is
\begin{equation}
B(Z \rightarrow l \bar l Z') \sim 2.1 \times 10^{-6} .
\end{equation}
The ratios of different final particles can be
easily obtained from the leptonic branching ratios of the $Z$, and the $Z'$
\begin{equation}
2\mu^+2\mu^- :\mu^+\mu^-\nu\bar \nu :
 2\nu 2\bar \nu : \tau \bar \tau + any = 7.5:42:33:18.
\end{equation}
The most interesting channel is a $Z'$ produced from neutrinos decays to
$\mu^+ \mu^-$ which has a probability of $35 \%$. The signal is characterized
by
a $\mu^+\mu^-$ pair with a fixed invariant mass which is smaller than $m_Z$.
The
residual energy is all missing.

In conclusion, when taking the L3 events seriously,
we feel that both $l^+l^-$ and $\gamma\gamma$ of the
$l^+l^-\gamma\gamma$ events need new physics to explain.

The authors would like to thank Han-qing Zheng at CERN
for drawing our attention to the new LEP events
as well as discussions, and Yue-kuan Li at KEK for assistance
in understanding the TRISTAN data, Tao Han and T. Kumita for reading
the primary draft of this paper and giving very useful comments.
The High Energy Physics Group of the Prairie View A \& M University is
operated by the Particle Detector Research Center, Texas National
Research Laboratory Commissions
under Contract No. RCFY 9208. Superconducting Super Collider Laboratory is
operated by the University Research Association, In., for the
U.S. Department of Energy under Contract No. DE-AC35-89ER40486.
The research of C.K. was supported in part by the U. S. Department of Energy
and in part by the Texas National Research Laboratory Commissions.

\newpage

\subsection*{Figure Captions}

\begin{enumerate}
\item The total branching ratio of the new events
 in Eq. (14) as a function of $m_{Z'}$ for
$m_{H_2} = 55$ GeV (dashed), 59.7 GeV (solid) and 65 GeV (dash-dotted).
\end{enumerate}

\end{document}